\documentclass[
 reprint,
 amsmath,amssymb,
 aps
]{revtex4-1}
\usepackage{bm}
\usepackage{subfigure}
\usepackage[dvipdfmx]{graphicx}
\usepackage{color}

\allowdisplaybreaks[1]

\begin{document}

\title{Dilepton production spectrum above $T_{\rm c}$ with a lattice quark propagator}
\author{Taekwang Kim}
 \email{kim@kern.phys.sci.osaka-u.ac.jp}
\author{Masayuki Asakawa}
 \email{yuki@phys.sci.osaka-u.ac.jp}
\author{Masakiyo Kitazawa}
 \email{kitazawa@phys.sci.osaka-u.ac.jp}
\affiliation{Department of Physics, Osaka University,
Toyonaka, Osaka 560-0043, Japan}
\date{\today}

\begin{abstract}
The dilepton production rate from the deconfined medium is 
analyzed with the photon self-energies constructed from
quark propagators obtained by lattice numerical simulation
for two values of temperature $T=1.5T_{\rm c}$ and $3T_{\rm c}$
above the critical temperature $T_{\rm c}$. 
The photon self-energy is calculated by the Schwinger-Dyson 
equation with the lattice quark propagtor and a vertex function 
determined so as to satisfy the Ward-Takahashi identity.
The obtained dilepton production rate at zero momentum 
exhibits divergences reflecting van Hove singularity,
and is significantly enhanced around $\omega\simeq T$
compared with the rate obtained by the perturbative analysis.
\end{abstract}

\pacs{Valid PACS appear here}

\maketitle 

\section{\label{sec:level1}INTRODUCTION}

Ultra-relativistic heavy ion collisions are the unique method to 
produce the
deconfined medium experimentally on the Earth \cite{k,HIC}.
Various observables are measured in the experiments \cite{k} 
to reveal properties of the deconfined medium and 
a variety of phenomena which come into play during the time evolution 
of the hot medium.
Among these observables, the dilepton production yield has a 
characteristic feature that the yield provides us direct 
signals from the primordial deconfined medium \cite{Rapp:chiral}, because 
dileptons once produced in the hot medium do not interact with 
and pass through the medium owing to their colorless nature.

The dilepton yield observed experimentally consists of the
sum of the dilepton production in each stage of the time 
evolution of the hot medium.
The dilepton production in heavy ion collisions is 
roughly classified into three processes
except for final state hadronic decays.
The first one is the hard process, in which dileptons are
produced by scatterings of hard partons in the colliding nuclei. 
The second and third ones are thermal radiations from the 
deconfined medium and confined medium, respectively.
The dileptons in low invariant mass region are usually expected
to be dominated by these thermal radiations, while 
dileptons from the hard process have 
relatively high transverse momenta and large invariant masses.
Experimental results on the dilepton production yield are 
usually compared with the baseline called cocktail,
which is the production yield estimated from the observed hadron 
abundances and their branching ratios into a dilepton pair.
If there is no dilepton production from hot medium,
dilepton production yield should be consistent with the cocktail
result.

At Relativistic Heavy Ion Collider (RHIC) at Brookhaven National 
Laboratory, $e^+e^-$ pair production yield is measured by two 
collaborations, STAR and PHENIX \cite{a,b}. 
Both of these collaborations reported that the pair production 
yield measured in Au+Au collisions at 
$\sqrt{s_{NN}}=200$ GeV has significant enhancements in the low 
invariant mass ($m$) region compared with the cocktail \cite{a,b,c},
although the descrepancy on the magnitude of the enhancement 
between the two experimental groups has not been settled down yet.
The result of PHENIX Collaboration shows that 
the yield at $m\simeq 500$ MeV is about one order larger than 
the cocktail \cite{a}.

When one performs an estimate of the dilepton production yield, 
one first calculates the dilepton production {\it rate} 
per unit time and unit volume from a static medium.
The dilepton production {\it yield} is then given by the
space-time integral of the rate from each volume element of the medium.
The dilepton production rate of a static medium is proportional to 
the imaginary part of virtual photon self-energy \cite{f,g,h}.
When the temperature ($T$) is asymptotically high, 
the photon self-energy can be calculated perturbatively.
Using the hard thermal loop (HTL) resummed perturbation theory \cite{d,e},
the dilepton production rate was calculated 
in Ref.~\cite{i} for lepton pairs with zero total three-momentum, 
and the result was extended in Ref.~\cite{j} to nonzero momentum.
It is, however,  nontrivial whether or not such perturbative analyses
well describe the production rate from the deconfined medium 
near the critical temperature $T_{\rm c}$, which has turned 
out to be a strongly-coupled system \cite{k}.
Moreover, it is known that the perturbative analyses in 
Refs.~\cite{i,j} are modified by proper inclusion of higher-order 
terms \cite{l}.
The analysis of higher order terms, however, is complicated
and it is still under debate whether the scheme is valid
for the whole kinetic region \cite{l,Laine:2013vma}.
For the description of the dilepton production rate 
in the deconfined phase near $T_{\rm c}$, therefore,
it is desirable to evaluate the rate 
without resort to perturbative methods.
In particular, the large enhancement observed at PHENIX \cite{a}
suggests a possibility that the dilepton production from the 
strongly-coupled medium above $T_{\rm c}$ has a large enhancement
compared with the perturbative results used in the previous 
analyses \cite{Rapp}.

There are several attempts of non-perturbative analyses 
for the dilepton production rate 
with lattice QCD \cite{Karsch:2001uw,Ding}.
When one investigates the rate on the lattice one must
take an analytic continuation from the imaginary-time correlator
computable on the lattice to the real-time photon self-energy.
This procedure, however, is an ill-posed problem, 
since the information on the imaginary-time correlators for 
discrete imaginary-time points obtained on the lattice is insufficient 
to reconstruct the continuous real-time function by itself \cite{MEM}.
In Ref.~\cite{Ding}, an ansatz for the spectral function
is introduced to avoid this problem.
An alternative way is to use the Baysian analysis such as the 
maximum entropy method \cite{MEM}.
The lattice correlator, however, is insensitive to the structure of
the spectrum in the low energy region \cite{app-n}.
The estimate of the low-energy spectrum on the lattice, therefore,
is a difficult problem with the presently existing methodologies.

The exact non-perturbative photon self-energy can be calculated 
by the Schwinger-Dyson equation (SDE) if we have the full quark 
propagator and the photon-quark vertex function.
Recently, an analysis of the non-perturbative quark propagator 
above $T_{\rm c}$ was performed on the lattice in the quenched approximation 
in Landau gauge \cite{m,n,o}.
In this series of analyses, 
the quark propagator was analyzed 
with the two-pole ansatz for the real-time propagator 
in order to carry out the analytic continuation.
This pole ansatz was employed motivated by 
the study of fermion propagator at nonzero temperature 
\cite{Baym:1992eu,Kitazawa:2007ep}.
It has been shown \cite{m,n,o} that this simple ansatz can 
reproduce the quark correlator obtained on the lattice 
over a rather wide range of bare quark mass and momentum.
It is therefore expected that
the obtained quark propagator well grasps the gist of
the non-perturbative nature of the quark propagator.

The purpose of the present study is to analyze the dilepton 
production rate using this quark propagator.
We construct the SDE with the quark propagators
obtained on the lattice in Ref.~\cite{o} 
and with the vertex function constructed so as to satisfy the 
Ward-Takahashi identity (WTI).
Our formalism, therefore, fulfills the conservation law of 
electric current.
In this analysis, we show that the obtained dilepton 
production rate exhibits an enhancement of one order or more
compared with the one from the free quark gas.
Compared with the perturbative result in Ref.~\cite{i},
our result has a qualitatively similar behavior at low 
$m$ region, while it exhibits an enhancement 
around $m\simeq T$ owing to van Hove singularity.
The effect of the vetex correction is also discussed in detail.

The outline of this paper is as follows. 
In the next section we introduce the SDE for the photon
self-energy and its components, lattice quark propagator and 
a veterx function satisfying the WTI. 
In Sec.~\ref{sec:rate}, we then solve the SDE
and present the form of the dilepton production rate in our formalism.
The rate without the vertex correction is also calculated in this section.
We then present the numerical result in Sec.~\ref{sec:results}.
The final section is devoted to a short summary.

\section{Schwinger-Dyson equation for photon self-energy}
\label{sec:level2}

\subsection{Schwinger-Dyson equation}

\begin{figure}
\begin{center}
\includegraphics[scale=0.7]{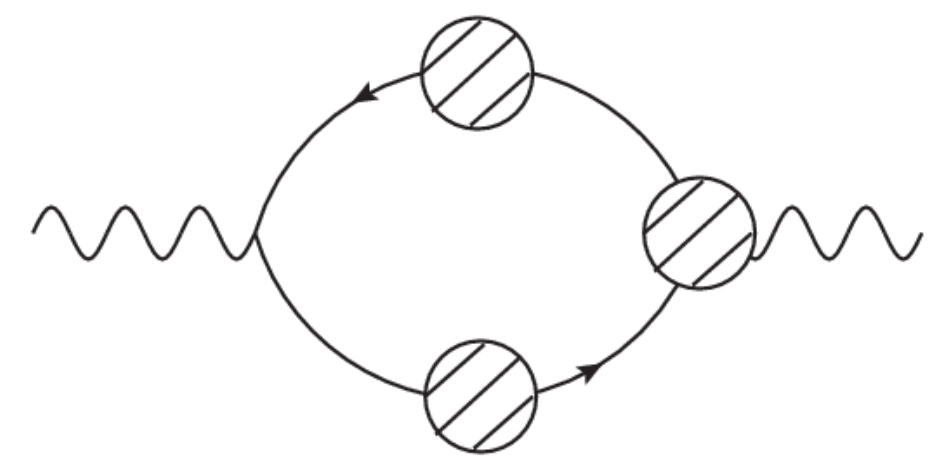}
\end{center}
\caption{
Diagramatic representation of the Schwinger-Dyson equation for 
the photon self-energy, Eq.~(\ref{eq:SDE}). 
The shaded circles represent the full quark propagator and 
the full vertex function.}
\label{fig:SDE}
\end{figure}

As dileptons are emitted from decays of virtural photons, 
the dilepton production rate from a medium per unit time and unit 
volume is related to the retarded self-energy 
$\Pi^R_{\mu\nu}(\omega,\bm{q})$ of the virtual photon as
\cite{f,g,h}
\begin{equation}
  \frac{d\Gamma}{d\omega d^3q}
  =\frac{\alpha}{12\pi^4}\frac{1}{Q^2}
  \frac{1}{\mbox{e}^{\beta \omega}-1}
  \mbox{Im}\Pi_\mu^{{R},\mu}(\omega,\bm{q})
  \label{eq:rate}
\end{equation}
at the leading order of fine structure constant $\alpha$
with $Q^2 = \omega^2-\bm{q}^2$ 
and the inverse temperature $\beta=1/T$.
With the SDE in Matsubara formalism,
the exact photon self-energy is given by the full quark propagator
$S(P)$ and the full photon-quark vertex $\Gamma_\mu(P+Q,P)$ as
\begin{align}
\Pi_{\mu\nu}\left(i\omega_m,\bm{q}\right)  = & - \sum_{\rm f} e_{\rm f}^2 T\sum_{n}\int\frac{d^3p}{\left(2\pi\right)^3} \nonumber \\
\  \mathrm{Tr}_{{\rm C}} \mathrm{Tr}_{{\rm D}} & \left[S(P)\gamma_{\mu}S(P+Q)
\Gamma_\nu(P+Q,P)\right], \label{eq:SDE} 
\end{align}
where $\omega_m=2\pi Tm$ and $\nu_n=(2n+1)\pi T$ 
with integers $m$ and $n$ are the Matsubara frequencies 
for bosons and fermions, respectively, 
$P_\mu = (i\nu_n,\bm{p})$ is the four-momentum of quarks,
and $e_{\rm f}$ is the electric charge of a quark with an index 
${\rm f}$ representing the quark flavor. 
The color, flavor, and Dirac indices of $S(P)$ are suppressed
for notational simplicity.
Tr$_{\rm C}$ and Tr$_{\rm D}$ denote the trace over color 
and Dirac indices, respectively.
We note that since we take Landau gauge in this calculation,
off-diagonal elements in color space disappear. As a result,
the trace over
the color indices gives a factor $3$ in Eq.~(\ref{eq:SDE}).
Equation~(\ref{eq:SDE}) is graphically shown in Fig.~\ref{fig:SDE},
in which the shaded circles represent the full propagator and 
vertex function.
The retarded photon self-energy is obtained by the analytic continuation,
\begin{eqnarray}
\Pi^R_{\mu\nu}(\omega,\bm{q}) 
& = & 
\Pi_{\mu\nu}(i\omega_m,\bm{q})|_{i\omega_m \rightarrow \omega+i\eta}.
\end{eqnarray}
In the following, we consider the two-flavor system 
with degenerate $u$ and $d$ quarks, in which $\sum_{\rm f} e_{\rm f}^2 = 5e^2/9$.
In this study we also limit our attention to the $\bm{q}=0$ case.

\subsection{Lattice Quark Propagator and Spectral Function}
\label{sec:S}

In the present study, we use the quark propagator obtained 
on the quenched lattice in Ref.~\cite{o} as the full quark 
propagator in Eq.~(\ref{eq:SDE}).
In this subsection, after a brief review on the general 
property of the quark propagator we describe how to implement
the results in Ref.~\cite{o} in our analysis.

On the lattice with a gauge fixing, one can measure
the imaginary-time quark propagator,
\begin{equation}
  S_{\mu\nu}(\tau,\bm{p}) 
  = \int d^3x \mbox{e}^{-i\bm{p}\cdot\bm{x}}
  \langle 
  \psi_\mu(\tau,\bm{x})\bar{\psi}_\nu(0,\bm{0})\rangle,
\label{eq:S(tau)}
\end{equation}
where $\psi_\mu(\tau,\bm{x})$ is the quark field 
with the Dirac index $\mu$. 
Here, $\tau$ is the imaginary time restricted to the interval 
$0 \leq \tau < \beta$.
For the moment, the Dirac indices $\mu$ and $\nu$ 
of the quark propagator are explicitly shown.
The Fourier transform of the quark correlator, 
\begin{align}
  S_{\mu\nu}(i\nu_n,\bm{p})
  = \int_0^\beta d\tau  
  {\rm e}^{i\nu_n\tau} S_{\mu\nu}(\tau,\bm{p}),
\label{eq:S(nu)}
\end{align}
is written in the spectral representation as 
\begin{equation}
  S_{\mu\nu}(i\nu_n,\bm{p}) 
  = -\int_{-\infty}^\infty d\nu'\frac{\rho_{\mu\nu}(\nu',\bm{p})}{\nu'-i\nu_n}
  \label{eq:S}
\end{equation}
with the quark spectral function $\rho_{\mu\nu}(\nu',\bm{p})$.
The spectral function is related to the imaginary-time correlator
Eq.~(\ref{eq:S(tau)}) as
\begin{equation}
S_{\mu\nu}\left(\tau,\bm{p}\right) = \int_{-\infty}^{\infty}d\nu\frac{\mbox{e}^{\left(1/2-\tau/\beta\right)\beta \nu}}{\mbox{e}^{\beta \nu/2}+\mbox{e}^{-\beta \nu/2}}\rho_{\mu\nu}\left(\nu,\bm{p}\right).
\label{eq:d}
\end{equation}

In the deconfined phase in which the chiral symmetry is restored, 
the quark propagator anticommutes with $\gamma_5$. 
In this case, the spectral function can be decomposed with the projection operators $\Lambda_{\pm}\left(\bm{p}\right)=\left(1\pm\gamma_0\hat{\bm{p}}\cdot\bm{\gamma}\right)/2$ as
\begin{equation}
  \rho(\nu,\bm{p}) 
  = \rho_+(\nu,p) \Lambda_+(\bm{p}) \gamma_0 
  + \rho_-(\nu,p) \Lambda_-(\bm{p}) \gamma_0,
\end{equation}
with $p=|\bm{p}|$, $\hat{\bm{p}}=\bm{p}/p$, and 
\begin{equation}
  \rho_\pm(\nu,p) 
  = \frac12 \mbox{Tr}_{\rm D}\left[\rho\left(\nu,\bm{p}\right)\gamma_0\Lambda_\pm\left(\bm{p}\right)\right].
\end{equation}
It is shown from the anticommutation relations of $\psi$ and $\bar\psi$
that the decomposed spectral functions satisfy the sum rules,
\begin{equation}
  \int d\nu\rho_\pm(\nu,p) = 1.
  \label{eq:sumrule}
\end{equation}
Using charge conjugation symmetry, one can show that 
$\rho_\pm(\nu,\bm{p})$ satisfy 
\cite{o}
\begin{equation}
  \rho_\pm(\nu,p) = \rho_\mp(-\nu,p).
\end{equation}

On the lattice, 
one can measure the imaginary-time correlator Eq.~(\ref{eq:S(tau)})
for discrete imaginary times.
To obtain the quark propagator one has to deduce the spectral 
function from this information.
In Refs.~\cite{m,n,o}, the quark correlator in Landau gauge 
is analyzed on the lattice with the quenched approximation,
and the quark spectral function is analyzed 
with the two-pole ansatz,
\begin{equation}
  \rho_+(\nu,p)
  = Z_+(p) \delta\left(\nu-\nu_+(p)\right)
  + Z_-(p) \delta\left(\nu+\nu_-(p)\right),
  \label{eq:2pole}
\end{equation}
where $Z_\pm(p)$ and $\nu_\pm(p)$ are the residues and 
positions of the poles, respectively. 
The four parameters, $Z_\pm(p)$ and $\nu_\pm(p)$, are determined
by fitting
the correlators obtained on the lattice for each $p$.
The two poles in Eq.~(\ref{eq:2pole}) at $\nu_+(p)$ and $\nu_-(p)$,
respectively, correspond to the normal and plasmino modes in 
the HTL approximation.
In fact, the study of the momentum and bare quark mass, $m_0$, 
dependences of the fitting paramters
\cite{m,n,o}  shows that the behavior of these parameters is
consistent with this observation \cite{Baym:1992eu,o2};
for example, for large $m_0$ or $p$ the residue of the plasmino
mode $Z_-(p)$ becomes small and the propagator approaches 
that of the free quark.
The restoration of the chiral symmetry for massless quarks
above $T_{\rm c}$ is also checked explicitly on the lattice by 
measuring of the scalar term in the massless quark propagator 
\cite{n,o}.

\begin{figure}
\begin{center}
\includegraphics{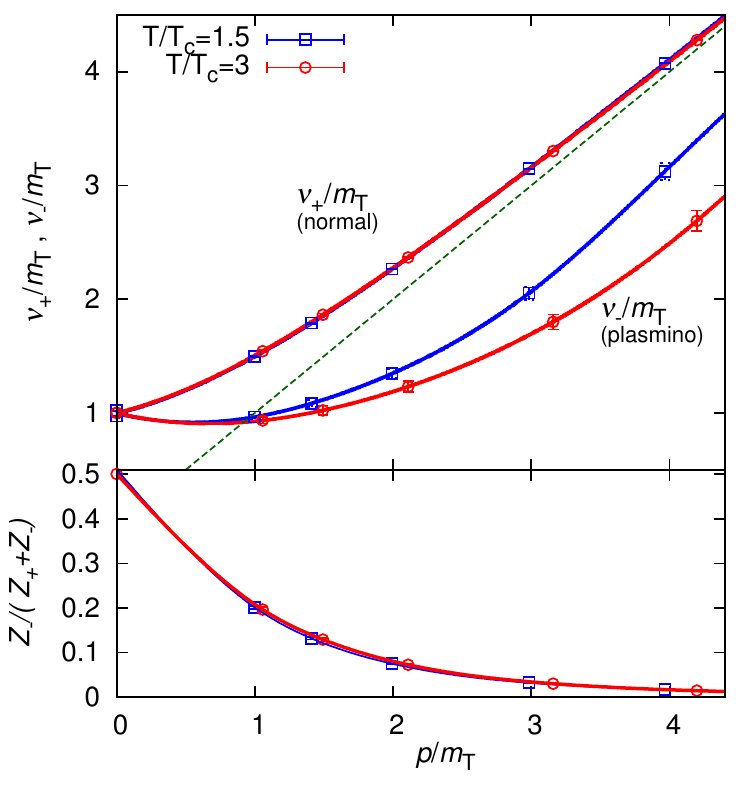}
\end{center}
\caption{
Open sympols show the momentum dependence of the parameters 
$\nu_+(p)$, $\nu_-(p)$, and $Z_-(p)/(Z_+(p)+Z_-(p))$ obtained 
on the lattice in Ref.~\cite{o}.
The solid lines represent their interpolation obtained 
by the cubic spline method.
The dashed line represents the light cone.}
\label{fig:one}
\end{figure}

In Fig.~\ref{fig:one}, we show the fitting result of each parameter
in Eq.~(\ref{eq:2pole}) for massless quarks as a function of $p$ 
for $T=1.5T_{\rm c}$ and $3T_{\rm c}$ obtained in Ref.~\cite{o}.
These analyses are performed on the lattice with the volume $128^3\times16$,
where both the lattice spacing and finite volume effects 
are found to be small \cite{o}.
In the upper panel, $p$ dependences of $\nu_\pm(p)$, i.e. the dispersion
relations of the normal and plasmino modes, are shown by the open symbols.
The vertical and horizontal axes are normalized by the thermal mass 
$m_{\rm T}$ defined by the value of $\nu_\pm(p)$ at $p=0$:
The value of $m_{\rm T}$ obtained on the lattice after the extrapolation 
to the infinite volume limit is $m_{\rm T}/T=0.768(11)$ and $0.725(14)$ 
at $T=1.5T_{\rm c}$ and $3T_{\rm c}$, respectively \cite{o}.
The lower panel shows the relative weight of the plasmino residue,
$Z_-/(Z_++Z_-)$.
This figure shows that the weight becomes smaller as $p$ increases,
which indicates that the quark propagator for large $p/T$ is 
dominated by the normal mode.
A result similar to that shown in Fig.~\ref{fig:one} is obtained
by the Schwinger-Dyson approach for the quark propagator 
\cite{Mueller:2010ah}.

Although the lattice data are available only for discrete 
values of $p$, we must have the quark propagator as a 
continuous function of $p$ to solve the SDE.
For this purpose, we take the interpolation and extrapolation
of the lattice data by the cubic spline method.
From the charge conjugation symmetry one can show that 
${\rm d}\nu_+(p)/{\rm d}p=-{\rm d}\nu_-(p)/{\rm d}p$, 
${\rm d}^2\nu_+(p)/{\rm d}p^2={\rm d}^2\nu_-(p)/{\rm d}p^2$, 
and $Z_+(p)=Z_-(p)$ for $p=0$ \cite{m,o}.
These properties are taken into account in our 
cubic spline interpolation.
The lattice data are available only in the momentum range 
$p/T \lesssim 4.7$.
To take extrapolations to higher momenta, 
we extrapolate the parameters using an exponentially damping form 
for $Z_-/(Z_++Z_-)$, 
\begin{align}
  Z_-/(Z_++Z_-) = R{\rm e}^{-\alpha p},
  \label{eq:extrapolate1}
\end{align}
and $\nu_\pm(p)$ are extrapolated by functions,
\begin{align}
  \nu_\pm(p) = p + \beta^\pm_1 {\rm e}^{-\beta^\pm_2 p},
  \label{eq:extrapolate2}
\end{align}
which exponentially approach the light cone for large $p$. 
The parameters $R$, $\alpha$, 
and $\beta^\pm_i$ are determined 
in the cubic spline analysis.
The $p$ dependence of each parameter determined in 
this way is shown by the solid lines in Fig.~\ref{fig:one}.
We tested another extrapolation form by a polynomial,
$\nu_\pm(p) = p + \beta'^\pm_1/p + \beta'^\pm_2/p^2 + \cdots$,
but found that it hardly changes the dispersion relation.
Finally, we fix 
\begin{align}
Z_+ + Z_- = 1
\end{align}
throughout this paper to satisfy
the sum rule Eq.~(\ref{eq:sumrule}).

With the two-pole form of the spectral function
Eq.~(\ref{eq:2pole}), the quark propagator reads
\begin{align}
  S(i\nu_n,\bm{p}) &= S_+(i\nu_n,p) \Lambda_+ (\bm{p}) \gamma_0
  + S_-(i\nu_n,p) \Lambda_-(\bm{p}) \gamma_0,
  \nonumber \\
  &= \sum_{s=\pm} S_s(i\nu_n,p) \Lambda_s (\bm{p}) \gamma_0 ,
  \label{eq:S_lat}
\end{align}
where 
\begin{align}
  S_s (i\nu_n,p)  =  \frac{ Z_+(p) }{ i\nu_n - s \nu_+(p) }
  + \frac{ Z_-(p) }{ i\nu_n + s \nu_-(p) }. \label{eq:S_s}
\end{align}
The symbols $s=\pm$ on the right-hand side
are understood as the numbers $\pm1$.
Correspondingly, the inverse propagator is given by 
\begin{equation}
  S^{-1}(i\nu_n,\bm{p})
  = \sum_{s=\pm} S_s^{-1}(i\nu_n,p) \gamma_0  \Lambda_s(\bm{p}),
  \label{eq:Sinv_lat}
\end{equation}
with
\begin{eqnarray}
  S_s^{-1} (i\nu_n,p)
  & = & \frac{ (i\nu_n - s \nu_+(p))(i\nu_n + s \nu_-(p))}{i\nu_n - s E(p)},
  \label{eq:partofSinv_lat}
\end{eqnarray}
and 
\begin{align}
E(p)=-Z_+(p)\nu_-(p) + Z_-(p)\nu_+(p).
\label{ep}
\end{align}
Note that the inverse propagator has poles at $i\nu_n=\pm E(p)$.
These poles inevitably appear in the multipole ansatz, because 
the propagator Eq.~(\ref{eq:S_s}) has one zero point 
in the range of $\omega$ surrounded by two poles.
The form of the inverse propagator Eq.~(\ref{eq:partofSinv_lat}) 
will be used in the construction of the vetex function.
We will see that the poles at $i\nu_n=\pm E(p)$ give rise to 
additional terms in the dilepton production rate.

\subsection{Vertex Function}
\label{sec:vertex}

The SDE, Eq.~(\ref{eq:SDE}), requires the full photon-quark vertex
$\Gamma_\mu(P+Q,P)$ besides the full quark propagator.
So far, the evaluation of $\Gamma_\mu(P+Q,P)$ on the lattice at 
nonzero temperature has not been performed to the best of the 
authors' knowledge.
In the present study, we construct the vertex function 
from the lattice quark propagator respecting the Ward-Takahashi 
identity (WTI) as follows.

The gauge invariance requires that the vertex function must fulfill 
the WTI 
\begin{equation}
 Q_\mu\Gamma^\mu(P+Q,P) = S^{-1}(P+Q) - S^{-1}(P),
 \label{eq:WTI}
\end{equation}
with the inverse quark propagator $S^{-1}(P)$.
For $\bm{q}=0$, the temporal component $\Gamma_0$ is completely 
determined only by this constraint as follows.
First, in this case $\bm{q}\cdot\bm{\Gamma}$ should vanish
provided that $\Gamma_i$ ($i=1,2,3$) are not singular at $\bm{q}=0$.
Then, by substituting $\bm{q}\cdot\bm{\Gamma}=0$ 
in Eq.~(\ref{eq:WTI}) one obtains
\begin{eqnarray}
  \lefteqn{\Gamma_0(i\omega_m+i\nu_n,\bm{p};i\nu_n,\bm{p}) } \nonumber \\
  &= \frac{1}{i\omega_m}\left[S^{-1}(i\omega_m+i\nu_n,\bm{p}) 
    - S^{-1}(i\nu_n,\bm{p})\right].
  \label{eq:Gamma_0}
\end{eqnarray}

On the other hand, the spatial components $\Gamma_i$ cannot be
determined only with Eq.~(\ref{eq:WTI}) \cite{p}.
In the present study, we employ an approximation to neglect
the $\bm{q}$ dependence of 
$\Gamma_0(i\omega_m+i\nu_n,\bm{p}+\bm{q};i\nu_n,\bm{p})$
at $\bm{q}=0$. In other words we assume that 
\begin{align}
\partial \Gamma_0(i\omega_m+i\nu_n,\bm{p}+\bm{q};
i\nu_n,\bm{p})/\partial q_i |_{\bm{q}=0}=0.
\label{eq:dGamma_0/dq}
\end{align}
With this approximaiton and Eq.~(\ref{eq:WTI}), one obtains
\begin{align}
  \lefteqn{ q^i\Gamma_i(i\omega_m+i\nu_n,\bm{p}+\bm{q};i\nu_n,\bm{p}) }
  \nonumber \\
  & =  S^{-1}(i\omega_m+i\nu_n,\bm{p}+\bm{q})
  - S^{-1}(i\omega_m+i\nu_n,\bm{p}) .
\end{align}
By taking the leading-order terms in $\bm{q}$ on the both sides,
one has 
\begin{align}
  \lefteqn{ \Gamma_i(i\omega_m+i\nu_n,\bm{p}; i\nu_n,\bm{p}) }
  \nonumber \\
  &= \frac{\partial S^{-1}}{\partial p^i}(i\omega_m+i\nu_n,\bm{p})
  \nonumber \\
  &= \sum_{s=\pm} \frac{\partial S_s^{-1}(i\omega_m+i\nu_n,p)}{\partial p^i}
  \gamma_0 \Lambda_s(\bm{p})
  \nonumber \\
  & +  \sum_{s=\pm} S_s^{-1}(i\omega_m+i\nu_n,p) \gamma_0 
  \frac{\partial \Lambda_s(\bm{p})}{\partial p^i} ,
  \label{eq:Gamma_i}
\end{align}
where in the second equality, we used Eq.~(\ref{eq:Sinv_lat}).

We note that there is no a priori justification of 
Eq.~(\ref{eq:dGamma_0/dq}).
By expanding $\Gamma_0$ with respect to $\bm{q}$ at $\bm{q}=0$,
one obtains
\begin{eqnarray}
  \lefteqn{ i\omega_m\Gamma_0(i\omega_m+i\nu_n,\bm{p}+\bm{q};i\nu_n,\bm{p})}
  \nonumber \\
  &= &S^{-1}(i\omega_m+i\nu_n,\bm{p}) - S^{-1}(i\nu_n,\bm{p}) 
  \nonumber \\
  &\ &+ \bm{q}\cdot\bm{p}\gamma_0A(i\omega_m+i\nu_n,i\nu_n,\bm{p}^2)
  \nonumber \\
  &\ &+ (\bm{q}\cdot\bm{p})(\hat{\bm{p}}\cdot\bm{\gamma})
  B(i\omega_m+i\nu_n,i\nu_n,\bm{p}^2) 
  \nonumber \\
  &\ &+ \bm{q}\cdot\bm{\gamma}C(i\omega_m+i\nu_n,i\nu_n,\bm{p}^2)
  +O(\bm{q}^2),
\label{eq:ABC}
\end{eqnarray}
where $A$, $B$, and $C$ are unknown functions.
Our approximation corresponds to neglecting these functions.
Although these functions do not affect Eq.~(\ref{eq:Gamma_0})
at $\bm{q}=0$, the corresponding terms appear 
in Eq.~(\ref{eq:Gamma_i}) when these functions are nonzero.
The determination of the non-perturbative form of the 
photon-quark and gluon-quark vertices is generally difficult, 
and various approximations have been employed in studies 
of the SDE \cite{p,Harada,Gao,Fischer}.
It should be emphasized that the vertex functions,
Eqs.~(\ref{eq:Gamma_0}) and (\ref{eq:Gamma_i}), satisfy
the WTI and thus are advantageous
in light of the gauge invariance among various ans\"atze on the vertex
function. $\Gamma_0$, Eq.~(\ref{eq:Gamma_0}), is the same as that obtained
in Ref.~\cite{p} since it is uniquely determined
only from the WTI. 
On the other hand, $\Gamma_i$ differ from the ones in Ref.~\cite{p},
even when only the longitudinal part in Ref.~\cite{p} is concerned.
Introduction of the functions given in Eq.~(\ref{eq:ABC}) fills 
this difference.

\section{Dilepton production rate}
\label{sec:rate}

The goal of the present study is to obtain the dilepton production
rate with the lattice quark propagators and the vertex function
discussed in the previous section.
In this section, however, before the analysis of the full 
manipulation we first see the dilepton production rates in simpler 
cases: (1) free quark gas in Sec.~\ref{sec:rate1},
and (2) case with
the lattice quark propagators but with the bare vertex function 
in Sec.~\ref{sec:rate2}. 
The full analysis is then presented in Sec.~\ref{sec:rate3}.

\subsection{Free quark gas}
\label{sec:rate1}

The photon self-energy for the massless free quark gas 
is obtained by substituting the free quark propagator
$S(i\nu_n,\bm{p}) = 1/(i\nu_n\gamma_0-\bm{p}\cdot\bm{\gamma})$ 
and bare vertex function $\Gamma_\mu=\gamma_\mu$ into Eq.~(\ref{eq:SDE}).
The result of the dilepton production rate for the massless two-flavor 
case is given by \cite{o2} 
\begin{align}
  \left.\frac{d\Gamma}{d\omega d^3q}\right|_{\bm{q}=\bm{0}}=\frac{5\alpha^2}{36\pi^4} \left( f\left(\frac{\omega}{2}\right)\right)^2,
\label{eq:free}
\end{align}
where $f(\omega) = 1/( {\rm e}^{\beta\omega}+1 )$ is 
the Fermi distribution function.

\subsection{Dilepton production rate without vertex correction}
\label{sec:rate2}

Next, we calculate the photon self-energy and 
the dilepton production rate with the lattice quark propagators
Eq.~(\ref{eq:S_lat}) but with $\Gamma_\mu=\gamma_\mu$.
The photon self-energy obtained in this way, of course,
does not fulfill the gauge invariance.
The result obtained here, however, is helpful in understanding
the effect of the vertex correction, i.e. the role of the WTI.

The photon self-energy with the bare vertex is given by
\begin{align}
  \Pi_{\mu\nu}(i\omega_m,\bm{q})
  = & -\frac{5e^2}3 T\sum_{n}
  \int\frac{d^3p}{(2\pi)^3} \nonumber \\
  & \mathrm{Tr}_{\rm D} [S(i\nu_n,\bm{p})\gamma_{\mu}
    S (i\omega_m+i\nu_n,\bm{p}+\bm{q})\gamma_\nu]. 
  \label{eq:C2-1}
\end{align}
By substituting Eq.~(\ref{eq:S_lat}) into this formula,
we obtain 
\begin{align}
  \Pi_{\mu\nu}(i\omega_m,\bm{0})  
  =& -\frac{5e^2}3 T\sum_n \int \frac{d^3p}{\left(2\pi\right)^3}
  \nonumber \\
  & \sum_{s,t=\pm} S_s(i\nu_n,p) S_t(i\omega_m+i\nu_n,p)
  \nonumber \\
  & \times
  \mbox{Tr}_{\rm D} [\Lambda_s\gamma_0\gamma_\mu\Lambda_t\gamma_0\gamma_\nu].
  \label{eq:C2-2}
\end{align}
The trace in Eq.~(\ref{eq:C2-2}) is calculated with
\begin{align}
  \mbox{Tr}_{\rm D} [ \Lambda_s(\bm{p})\Lambda_t(\bm{p}) ] & = 2 \delta_{st},
  \label{eq:Tr_D1}
  \\
  \mbox{Tr}_D [ \Lambda_s(\bm{p})\gamma_0\gamma_i
  \Lambda_t(\bm{p})\gamma_0\gamma_i ] & = 2 \delta_{s,-t} + 2st 
  \hat{p}_i^2,
  \label{eq:Tr_D2}
\end{align}
where it is understood that
\begin{align}
  \delta_{++} = \delta_{--} & =1, \quad
  \delta_{+-} = \delta_{-+}=0, \\ 
  \delta_{s,-t}  = \delta_{s,\mp} & \quad ({\rm for~~}t=\pm).
\end{align}
Substituting Eqs.~(\ref{eq:Tr_D1}), (\ref{eq:Tr_D2}), and 
$\sum_i \hat{p}_i^2=1$ into Eq.~(\ref{eq:C2-2}), one obtains
\begin{align}
  \Pi_\mu^\mu(i\omega_m,\bm{0})  
  =& \frac{20e^2}3 T\sum_n \int \frac{d^3p}{\left(2\pi\right)^3}
  \nonumber \\
  & \sum_{s=\pm} S_s(i\nu_n,p) S_{-s}(i\omega_m+i\nu_n,p) ,
\end{align}
where 
\begin{align}
S_{-s}(i\nu_n+i\omega_m,p) & = S_{\mp}(i\nu_n+i\omega_m,p) \quad ({\rm for~~}s=\pm).
\end{align}

Using Eq.~(\ref{eq:S_s}) and taking the Matsubara sum 
and the analytic continuation $i\omega_m\to\omega+i\eta$, we obtain 
\begin{widetext}
\begin{align}
  \Pi_\mu^{R,\mu}(\omega,\bm{0}) =&
  \frac{-40\alpha}{3\pi}\int_{0}^{\infty} dp~p^2 
  \left\{ \frac{Z_+(p)^2 \left( 1-2f(\nu_+(p)) \right)}
    {\omega - 2\nu_+(p) + i\eta} \right. 
    + \frac{Z_-(p)^2 \left( 1-2f(\nu_-(p))\right)}
    {\omega - 2\nu_-(p) + i\eta}  
    \nonumber \\
    &\left. + \frac{2 Z_+(p) Z_-(p) \left(f(\nu_-(p))-f_+(\nu_+(p))\right)}
         {\omega-\nu_+(p)+\nu_-(p)+i\eta} \right\}.
\end{align}
Taking the imaginary part of this result,  we have 
\begin{align}
  \mbox{Im} \Pi_\mu^{R,\mu}(\omega,\bm{0}) =& 
  \frac{40\alpha}{3}\int_{0}^{\infty} dp~p^2 
  \left\{ Z_+(p)^2\delta(\omega-2\nu_+(p)) \left( 1-2f(\nu_+(p)) \right) \right.
  \nonumber \\
  &+ Z_-(p)^2\delta (\omega-2\nu_-(p)) \left( 1-2f(\nu_-(p)) \right)
  \nonumber \\
  &+ 2 \left. Z_+(p)Z_-(p) \delta (\omega-\nu_+(p)+\nu_-(p))
  \left( f(\nu_-(p))-f(\nu_+(p))\right) \right\} . 
  \label{eq:novert}
\end{align}

\end{widetext}

\begin{figure}[htbp]
\begin{center}
\subfigure[Annihilation]{
\includegraphics*[scale=0.6]{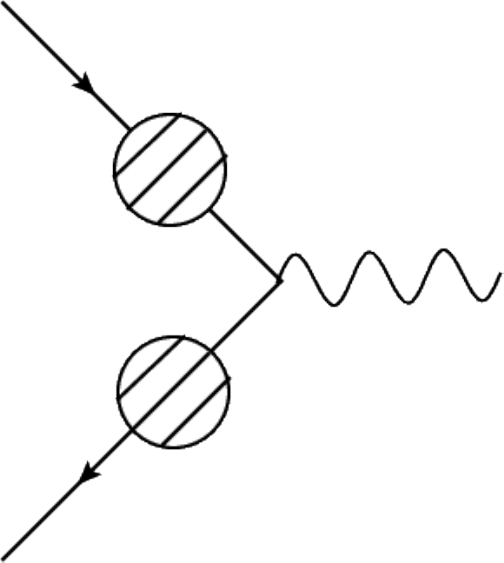}
\label{fig:annihilation2}}
\subfigure[Landau damping]{
\includegraphics*[scale=0.6]{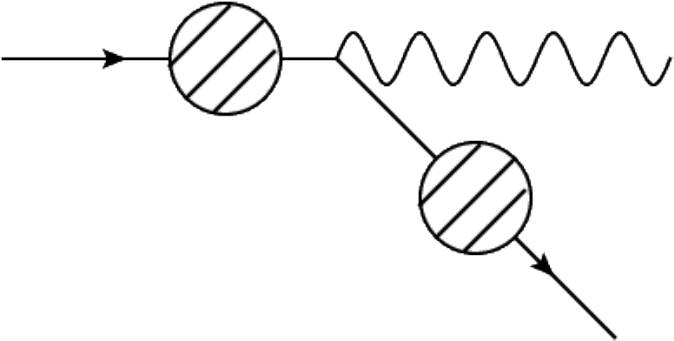}
\label{fig:Landau2}}
\end{center}
\caption{Photon production processes}
\label{fig:decay}
\end{figure}

The imaginary part of the photon self-energy is 
the difference between the annihilation and production rates
of virtual photons in medium.
The three terms in Eq.~(\ref{eq:novert}) represent 
different annihilation and production processes of a virtual 
photon.
The first and second terms in Eq.~(\ref{eq:novert}) represent 
the production of a virtural photon through the pair annihilation
of two normal modes and two plasmino modes, respectively, 
which are diagramatically shown in Fig.~\ref{fig:decay} (a),
and their inverse processes.
This can be checked from the arguments of the $\delta$-functions 
and thermal factors in Eq.~(\ref{eq:novert}).
The $\delta$-function in these terms represents the 
energy conservation during these processes, and 
the therrmal factor which is rewritten as 
\begin{align}
  1-2f(\omega) = \left( 1-f(\omega) \right)^2 - f(\omega)^2,
\end{align}
is the difference between the products of 
Pauli blocking effects and thermal distributions.
The existence of the residues $Z_\pm(p)$ in these terms 
is similarly understood.
The last term in Eq.~(\ref{eq:novert}) represents the Landau 
damping between normal and plasmino modes,
which is diagramatically shown in Fig.~\ref{fig:decay} (b).
Correspondingly, the thermal factor in this term can be rewritten as
\begin{align}
  f(\omega_1)-f(\omega_2) 
  = f(\omega_1) \left( 1-f(\omega_2) \right)
  - f(\omega_2) \left( 1-f(\omega_1) \right).
\end{align}

The $\delta$-functions in Eq.~(\ref{eq:novert}) can be integrated
out.
The form of the dilepton production rate after 
the integration over $p$ reads
\begin{widetext}
\begin{align}
  \mbox{Im}\Pi_\mu^{R,\mu}(\omega,\bm{0})
  =& \frac{40\alpha}{3}
  \left\{
    \frac{p^2Z_+(p)^2}{2|d\nu_+(p)/dp|}
    \left(1-2f(\nu_+(p))\right)
    \bigg|_{\omega=2\nu_+(p)} \right. 
    +\sum_l \frac{p_l^2Z_-(p_l)^2}
    {2|d\nu_-(p_l)/dp|}
    \left(1-2f(\nu_-(p_l))\right)
    \bigg|_{\omega=2\nu_-(p_l)} \nonumber \\
    &+ \sum_l \left.\frac{2p_l^2Z_+(p_l)Z_-(p_l)}
    {|d [ \nu_+(p_l) - \nu_-(p_l) ]/dp|}
    \left(f(\nu_-(p_l))-f(\nu_+(p_l))\right)
    \bigg|_{\omega=\nu_+(p_l)-\nu_-(p_l)}
    \right\},
  \label{eq:NoBC}
\end{align}
\end{widetext}
where the momentum $p$ in each term is given by 
the condition arising from the $\delta$-functions in 
Eq.~(\ref{eq:novert}).
Each term can take nonzero values only when
there exist momenta
satisfying this condition for a given $\omega$.
This gives the condition for $\omega$ at which each term takes
a nonzero value. For example, the first term takes nonzero values
for $\omega>2m_{\rm T}$.
Because the second and third terms can have multiple solutions
of $p$ for a fixed $\omega$, we represent this possibility by the
sum over $l$.
It is also notable that each term in Eq.~(\ref{eq:NoBC}) is 
inversely proportional to the derivative of
$\nu_\pm(p)$ and $\nu_+(p)-\nu_-(p)$;
they come from the density of states of each modes.
Accordingly, the dilepton production rate diverges when 
the derivatives vanish.
Such divergence is known as van Hove singularity.
In Sec.~\ref{sec:results}, we will see the appearance of 
such singularities in the dilepton spectrum.

\subsection{Dilepton production rate with vertex correction}
\label{sec:rate3}

Now, let us calculate the dilepton production rate with the 
lattice quark propagtor Eq.~(\ref{eq:S_lat}) and the full vertex 
functions Eqs.~(\ref{eq:Gamma_0}) and (\ref{eq:Gamma_i}).

When the full vertex function satisfying the WTI is used in 
Eq.~(\ref{eq:SDE}), the temporal component $\Pi_{00}$ for $\bm{q}=0$ 
vanishes.
One can easily check this explicitly by substituting
Eq.~(\ref{eq:Gamma_0}) into Eq.~(\ref{eq:SDE}).
For $\sum_{i=1}^{3}\Pi_{ii}$, by substituting Eqs.~(\ref{eq:S_lat}) 
and (\ref{eq:Gamma_i}) into Eq.~(\ref{eq:SDE}) one has
\begin{widetext}
\begin{align}
  \sum_{i=1}^{3} \Pi_{ii}(i\omega_m,\bm{0})  
  &= \frac{5e^2}{3}T\sum_n \int\frac{d^3p}{(2\pi)^3}
  \sum_{s,t,u=\pm}S_s(i\nu_n,p) S_t(i\nu_n+i\omega_m,p )
  \nonumber \\
  & \times \sum_{i=1}^{3} \left( \hat{p}_i 
  \frac{\partial S^{-1}_u(i\nu_n+i\omega_m,p)}{\partial p}
  \mbox{Tr}_{\rm D}[\Lambda_s\gamma_0\gamma_i\Lambda_t \Lambda_u]
  + \frac{ u S^{-1}_u(i\nu_n+i\omega_m,p) }{2p} 
  \mbox{Tr}_{\rm D}[\Lambda_s\gamma_0\gamma_i\Lambda_t\gamma_0
  (\gamma_i-(\hat{\bm{p}}\cdot\bm{\gamma})\hat{p}_i)] \right).
  \label{eq:Pi_full}
\end{align}
We substitute the following relations for the Dirac traces,
\begin{align}
  \mbox{Tr}_{\rm D}\left[\Lambda_s\gamma_0\gamma_i\Lambda_t \Lambda_s\right]
  &= 2s\hat{p}_i\delta_{st}\delta_{tu}, \\
  \mbox{Tr}_{\rm D}\left[\Lambda_s\gamma_0\gamma_i\Lambda_t \gamma_0\gamma_i\right]
  &= 2 \delta_{s,-t} + 2st\hat{p}_i^2, \\
  \mbox{Tr}_{\rm D}\left[\Lambda_s\gamma_0\gamma_i\Lambda_t
  \gamma_0(\hat{\bm{p}}\cdot\bm{\gamma})\right]
  &= 2\hat{p}_i\delta_{st},
  \label{eq:Tr_D3}
\end{align}
and obtain
\begin{eqnarray}
  \sum_i\Pi_{ii}(i\omega_m,\bm{0}) 
  = \frac{10e^2}{3}T\sum_n\int\frac{d^3p}{(2\pi)^3}
  \sum_{s=\pm} sS_s(i\nu_n,p) \left\{ \frac{\partial \mbox{ln}S^{-1}_s(i\nu_n+i\omega_m,p) }{\partial p}  
  - \frac1p
  \left( 1-\frac{S_{-s}(i\nu_n+i\omega_m,p)}{S_s(i\nu_n+i\omega_m,p)} \right)
  \right\}, 
  \label{eq:D-1}
\end{eqnarray}
\end{widetext}
where to obtain the first term we used
\begin{align}
  S_s(i\nu_n,p)
  \frac{\partial S_s^{-1}(i\nu_n,p)}
       {\partial p}
  = \frac{\partial \mbox{ln}
    S_s^{-1}(i\nu_n,p)}
  {\partial p}.
\end{align}

Up to now, the calculation relies only on the decomposition 
Eq.~(\ref{eq:S_lat}), which is valid for the chiral symmetric
quark propagator, and the form for the vertex Eq.~(\ref{eq:Gamma_i}).
The result Eq.~(\ref{eq:D-1}) thus is valid for any form of 
the quark propagator $S_\pm(i\nu_n,p)$.
Now, we use the two-pole form of the quark propagator, 
Eqs.~(\ref{eq:S_s}) and (\ref{eq:partofSinv_lat}).
The term including $\partial \ln S_s^{-1}/\partial p$ in 
Eq.~(\ref{eq:D-1}) is then calculated to be 
\begin{widetext}
\begin{align}
  \lefteqn{ T\sum_n sS_s(i\nu_n,p) \frac{\partial \mbox{ln}S^{-1}_s(i\nu_n+i\omega_m,\bm{p}) }{\partial p}  }
  \nonumber \\
  =& -T\sum_n \left(\frac{Z_+}{i\nu_n-s\nu_+}+\frac{Z_-}{i\nu_n+s\nu_-}\right) 
  \left(\frac{d\nu_+/dp}{i\nu_n+i\omega_m-s\nu_+}-\frac{d\nu_-/dp}{i\nu_n+i\omega_m+s\nu_-}-\frac{dE/dp}{i\nu_n+i\omega_m-sE}\right)
  \nonumber \\
  =& -\left( \frac{Z_+d\nu_-/dp}{i\omega_m+s\nu_++s\nu_- }
  + \frac{Z_-d\nu_+/dp}{i\omega_m-s\nu_+-s\nu_- } \right)
  \left( f(s\nu_+) + f(s\nu_-) -1 \right)
  \nonumber \\
  & - \frac{Z_+dE/dp}{ i\omega_m+s\nu_+-sE }
  \left( f(s\nu_+) - f(sE) \right)
  - \frac{Z_-dE/dp}{i\omega_m-s\nu_--sE }
  \left( f(s\nu_-) - f(sE) \right) ,
  \label{eq:D-3}
\end{align}
where the Matsubara sum over $n$ is taken in the last equality.
The remaining part of Eq.~(\ref{eq:D-1}) is calculated as follows:
\begin{align}
  \lefteqn{ T\sum_n sS_s(i\nu_n,p) 
  \left( 1-\frac{S_{-s}(i\nu_n+i\omega_m,p)}{S_s(i\nu_n+i\omega_m,p)} \right)
  \frac1p }
  \nonumber \\
  =& T\sum_n \left(\frac{Z_+}{i\nu_n-s\nu_+}+\frac{Z_-}{i\nu_n+s\nu_-}\right) 
  \left(\frac{F_1}{i\nu_n+i\omega_m-sE}
  +\frac{F_2}{i\nu_n+i\omega_m+s\nu_+}
  +\frac{F_3}{i\nu_n+i\omega_m-s\nu_-}\right)
  \\
  =& \frac{Z_+F_1}{i\omega_m+s\nu_+-sE}\left(f(sE)-f(s\nu_+)\right)
  +\frac{Z_-F_1}{i\omega_m-s\nu_--sE}\left(f(sE)-f(-s\nu_-)\right)
  \nonumber \\
  & +\frac{Z_+F_2}{i\omega_m+2s\nu_+}\left(f(-s\nu_+)-f(s\nu_+)\right)
  +\frac{Z_-F_2}{i\omega_m+s\nu_+-s\nu_-}\left(f(-s\nu_+)-f(-s\nu_-)\right)
  \nonumber \\
  & +\frac{Z_+F_3}{i\omega_m+s\nu_+-s\nu_-}\left(f(s\nu_-)-f(s\nu_+)\right)
  +\frac{Z_-F_3}{i\omega_m-2s\nu_-}\left(f(s\nu_-)-f(-s\nu_-)\right) , 
  \label{eq:D-4}
\end{align}
where 
\begin{eqnarray}
  F_1 &=& - \frac{ 2E(\nu_-+E)(\nu_+-E) }{ p(\nu_--E)(\nu_++E) },
  \nonumber \\
  F_2 &=& \frac{ 2\nu_+(\nu_+-E)(\nu_+-\nu_-) }{ p(\nu_++E)(\nu_++\nu_-) },
  \nonumber \\
  F_3 &=& \frac{ 2\nu_-(\nu_+-\nu_-)(\nu_-+E) }{ p(\nu_++\nu_-)(\nu_--E) }.
  \nonumber
\end{eqnarray}
Here, each combination of $\nu_\pm$ and $E$ in the parantheses is set 
to become positive; this can be checked by the relation 
$\nu_+>\nu_->-E>0$.

Combining these results, Eq.~(\ref{eq:D-1}) is calculated to be
\begin{eqnarray}
  \sum_{i=1}^3\Pi_{ii}\left(i\omega_m,\bm{0}\right) 
  & = & -\frac{10e^2}{3}\int\frac{d^3p}{\left(2\pi\right)^3}
  \sum_{s=\pm} s
  \left\{
  \frac{2Z_+^2\nu_+\bar{\Omega}}{p(\nu_++E)}
  \frac{1-2f(\nu_+) }{i\omega_m+2s\nu_+}
  +\frac{2Z_-^2\nu_-\bar{\Omega}}{p(\nu_--E)}
  \frac{ 1-2f(\nu_-) }{i\omega_m+2s\nu_-}
  \right. 
  \nonumber \\
  &\ &
  +2Z_+Z_-\bar{\Omega}\frac{\bar{\Omega}E-2\omega_+\omega_-}{p(\nu_++E)(\nu_--E)}
  \frac{ f(\nu_-)-f(\nu_+) }{i\omega_m-s\nu_++s\nu_-}
  \nonumber \\
  &\ &
  -\left(Z_+\frac{d\nu_-}{dp}-Z_-\frac{d\nu_+}{dp}\right)
  \frac{ 1-f(\nu_+)-f(\nu_-) }{i\omega_m+s\nu_+ +s\nu_- }
  \nonumber \\
  &\ & \left.
  + \left( - \frac{2Z_+Z_-E(\nu_+ + \nu_-)^2}{p(\nu_++E)(\nu_--E)} 
  - \frac{dE}{dp} \right)
  \left( Z_+ \frac{ f(E)-f(\nu_+) }{i\omega_m+s\nu_+-sE}
  + Z_- \frac{ f(-E)-f(\nu_-) }{i\omega_m+s\nu_-+sE} \right)
  \right\}
  \label{eq:D-5}
\end{eqnarray}
with $\bar{\Omega}=\nu_+-\nu_-$.

By taking the analytic continuation $i\omega_m\rightarrow \omega+i\eta$,
and taking the imaginary part, we obtain 
\begin{align}
  \mbox{Im}\Pi_\mu^{R,\mu}(\omega,\bm{0})
  = &
  -\frac{20\alpha}3 \int dp~p^2\bigg[
  -\frac{2\bar{\Omega}}{p} 
  \left\{ \frac{Z_+^2\nu_+}{\nu_++E} \left( 1-2f(\nu_+) \right)
  \delta(\omega-2\nu_+)
  + \frac{Z_-^2\nu_-}{\nu_--E} \left( 1-2f(\nu_-) \right)
  \delta( \omega-2\nu_- )\right.
  \nonumber \\
  &
  \left.-Z_+Z_-\frac{\bar{\Omega}E-2\nu_+\nu_-}{(\nu_++E)(\nu_--E)}
  \left( f(\nu_-)-f(\nu_+) \right) \delta(\omega-\nu_++\nu_-) \right\}
  \nonumber \\
  & +\left(Z_+\frac{d\nu_-}{dp}-Z_-\frac{d\nu_+}{dp}\right) \left( 1-f(\nu_+)-f(\nu_-) \right) \delta(\omega-\nu_+-\nu_-)
  \nonumber \\
  & +\left( \frac{2Z_+Z_-E(\nu_+ + \nu_-)^2}{p(\nu_++E)(\nu_--E)} + \frac{dE}{dp} \right)
  \nonumber \\
  & \times \left\{
  Z_+ \left(1-f(-E)-f(\nu_+) \right) \delta( \omega-\nu_++E )
  + Z_- \left(f(-E)-f(\nu_-) \right) \delta( \omega-\nu_--E )
  \right\} \bigg]
  \nonumber \\
  & + (\omega\to-\omega) .
\label{eq:full_rate}
\end{align}
\end{widetext}

Now let us inspect the physical meaning of each term 
in Eq.~(\ref{eq:full_rate}).
From the $\delta$-functions and thermal factors, 
one finds that the two terms in the first line 
represent the pair creation and annihilation processes 
of normal and plasmino modes, respectively.
The second line corresponds to the Landau damping.
These terms have corresponding counterparts in 
Eq.~(\ref{eq:novert}), although the coefficients of 
these terms are modified by the vertex correction.
The term in the third line in Eq.~(\ref{eq:full_rate})
can be interpreted as the pair annihilation and creation of 
a normal mode and a plasmino one.
This process does not appear in Eq.~(\ref{eq:novert}), 
and can manifest itself as a consequence of the vertex correction.
We note that the similar process exists in the formula obtained
in Ref.~\cite{i}.
In this way, the terms in the first three lines in 
Eq.~(\ref{eq:full_rate}) can be understood as the 
annihilation, creation, and scattering processes of quark quasi-particles.
We also note that the Landau damping of two normal or two plasmino
modes do not exist in Eq.~(\ref{eq:full_rate}), because such a process
can exist only for $\omega=0$ at $\bm{q}=0$.

On the other hand, one cannot give such interpretations to 
the terms in the fourth and fifth lines in Eq.~(\ref{eq:full_rate}).
From the $\delta$-functions and the thermal factors,
these terms seem to represent the decay and creation rates 
with a quasi-particle mode with energy $\pm E$, 
which, however, does not exist in the quark propagtor Eq.~(\ref{eq:S_s}).
Mathematically, these terms come from the poles in the vertex 
function Eq.~(\ref{eq:Gamma_i}).
The poles appear in the vertex function via the WTI,
Eq.~(\ref{eq:WTI}) and the fact that the analytic continuation of 
the propagator $S_s(i\nu_n,p)$ gives zero points at energies $\pm E$.
As discussed in Sec.~\ref{sec:S}, the zero in $S_s(\omega,p)$ 
inevitably appears between the two poles in the two-pole form of 
the quark propagator Eq.~(\ref{eq:S_s}).

Another remark on Eq.~(\ref{eq:full_rate}) is the 
sign of each term in Eq.~(\ref{eq:full_rate}).
In Eq.~(\ref{eq:full_rate}), all terms are separately 
positive definite for $\omega>0$ except for the one in the third line,
which becomes negative for sufficiently large $\omega$.
The negative contribution of this term is, however, canceled out 
by the last term; 
we have checked that the sum of these terms is always positive.
The total dilepton production rate for $\omega>0$ therefore 
takes a positive value as it should be.

\begin{figure}
\begin{center}
\includegraphics{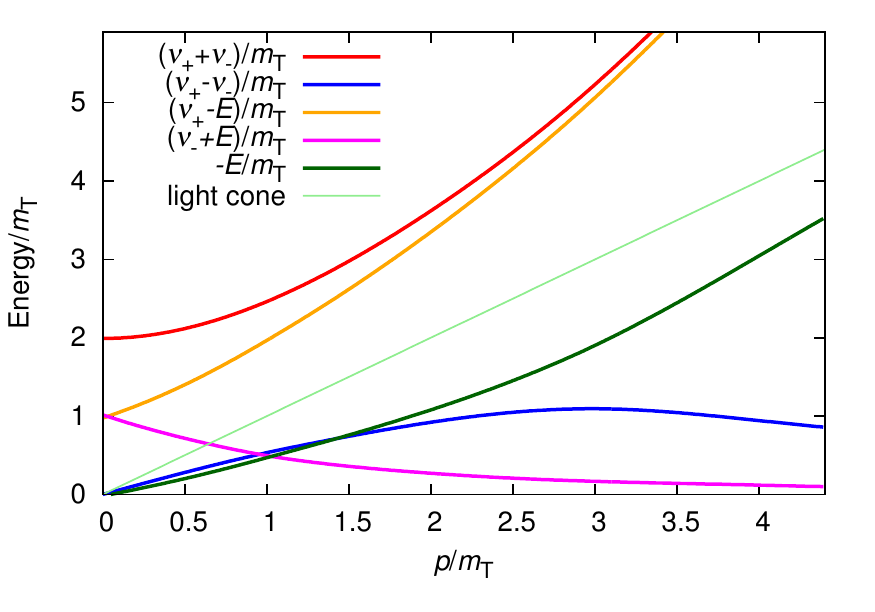}
\end{center}
\caption{
Momentum dependences of various functions composed of $\nu_\pm(p)$
and $E$.}
\label{fig:add_disp}
\end{figure}

In Fig.~\ref{fig:add_disp}, we show 
the dispersion relation of $E(p)$ for $T=1.5T_{\rm c}$.
For $p=0$, $E(p)$ vanishes because of chiral symmetry, 
while $-E(p)$ approaches $p$ for large $p$.
The figure shows that $-E(p)$ is a monotonically increasing
function of $p$.
The result for $T=3T_{\rm c}$ is qualitatively the same.
In Fig.~\ref{fig:add_disp}, we also show 
the combinations of the dispersion relations appearing in 
the $\delta$-functions in Eq.~(\ref{eq:full_rate}),
\begin{align}
\nu_++\nu_- , \quad
\nu_+-\nu_- , \quad
\nu_+-E , \quad
\nu_-+E.
\end{align}
From the figure, one sees that 
$\nu_+-E$ and $\nu_-+E$ are monotonically increasing and 
decreasing functions of $p$, respectively, starting from 
$m_{\rm T}$ at $p=0$.
Also, $\nu_+-E>m_{\rm T}$ and $0<\nu_-+E<m_{\rm T}$ are satisfied.
These behaviors become transparent by rewriting these combinations as
\begin{align}
\nu_+ - E &= Z_+ ( \nu_+ + \nu_- ),
\\
\nu_- + E &= Z_- ( \nu_+ + \nu_- ),
\end{align}
where we used Eq. (\ref{ep}).

We finally comment on the limiting behaviors of Eq.~(\ref{eq:full_rate}).
First, in our two-pole ansatz the quark propagator for 
massless free quarks is obtained by setting 
\begin{align}
Z_+(p)=1, \quad Z_-(p)=0, \quad \nu_+(p)=p.
\label{eq:Zfree}
\end{align}
Equation~(\ref{eq:full_rate}) thus should reproduce the 
photon self-energy of the free quark gas, 
when Eq.~(\ref{eq:Zfree}) is substituted.
This can be explicitly checked as follows.
By substituting $Z_-=0$, all terms including $Z_-$ vanishes.
Since $E=-\nu_-$ for $Z_-=0$, 
the third and fourth lines in Eq.~(\ref{eq:full_rate}) 
cancel out with each other 
without constraints on $\nu_-(p)$.
Only the first term in Eq.~(\ref{eq:full_rate}) thus survives,
which gives the free quark result.
Second, 
our result on the dilepton production rate approaches 
the free quark one in the large $\omega$ limit,
because the lattice quark propagator used in this study
reproduces Eq.~(\ref{eq:Zfree}) at large momentum.
This behavior will be explicitly checked in the next section.

\section{Numerical results}
\label{sec:results}

\begin{figure}
\begin{center}
\includegraphics{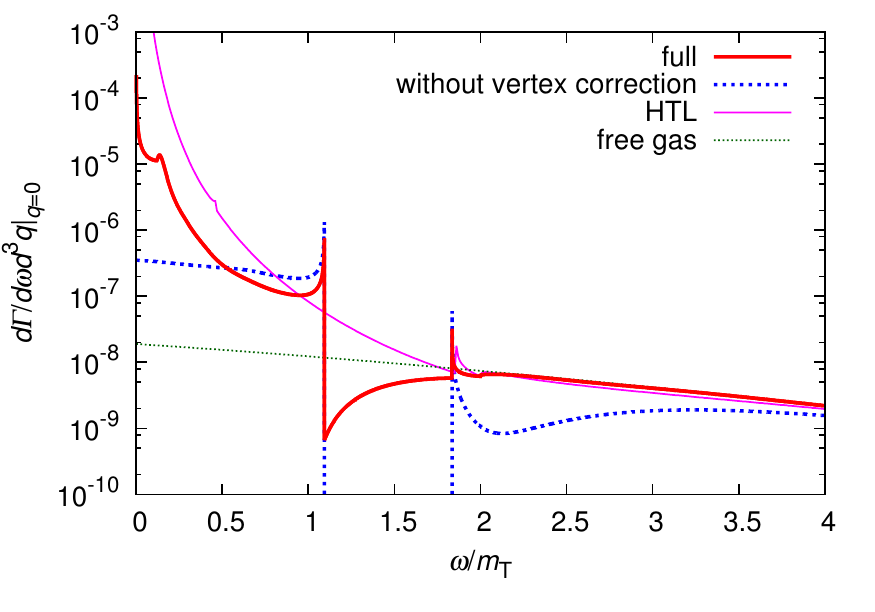}
\end{center}
\caption{
Dilepton production rate at zero momentum for $T=1.5T_{\rm c}$.
The result without vertex correction is also plotted.
Thin lines represent the HTL and free quark results.
}
\label{fig:two}
\end{figure}

Now let us see the numerical results on the dilepton production
rate obtained in the previous section.
In Fig.~\ref{fig:two}, we present the $\omega$ dependence of 
the dilepton production rate for $T=1.5T_{\rm c}$.
In the figure, we also plot 
the result without vertex correction, Eq.~(\ref{eq:novert}),
together with the rates obtained by the HTL calculation \cite{i} 
and the free quark gas, Eq.~(\ref{eq:free}).
The value of the thermal mass $m_{\rm T}$ is taken from the 
the result obtained on the lattice \cite{o}.

Figure~\ref{fig:two} shows that 
the production rate with the lattice quark propagators
has divergences at two energies, $\omega/m_{\rm T}=\omega_1/m_{\rm T}\simeq1.1$
and $\omega/m_{\rm T}=\omega_2/m_{\rm T}\simeq1.8$.
For $\omega < \omega_1$, our result, as a whole, behaves similarly to 
the HTL one \cite{i}, i.e. it increases as $\omega$ 
decreases, although our production rate is smaller than 
the perturbative one for small $\omega$.
Near $\omega_1$, however, 
it shows a prominent enhancement and exceeds the latter.
The region, where the large production rate is obtained, is 
located around $m_{\rm T}$.
Therefore, it can be possible that the production yield obtained 
by integrating this rate has the large enhancement below 
several hundred MeV,
where the enhancement in the experimentally-observed dilepton
spectra at RHIC \cite{a,b,c} exists.
The rate has a discontinuity at $\omega=\omega_1$,
and is significantly suppressed compared with Eq.~(\ref{eq:free})
for $\omega_1<\omega<\omega_2$.
The rate has another discontinuity at $\omega=\omega_2$, 
above which the rate is close to the free quark one.
In the dilepton rate without vertex correction, 
one also finds two divergences at $\omega=\omega_1$ and $\omega_2$,
while the rate vanishes for $\omega_1<\omega<\omega_2$.

\begin{figure}[htbp]
\begin{center}
\includegraphics{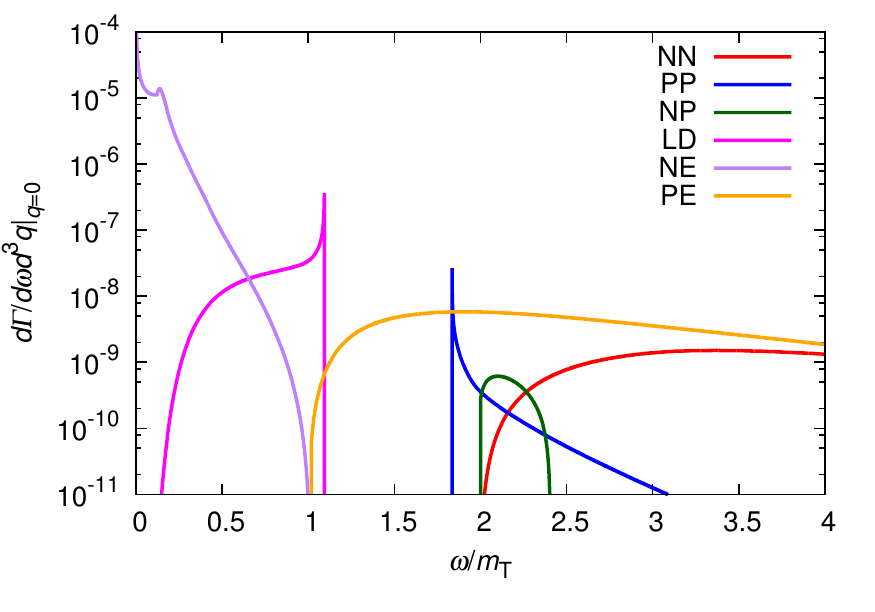}
\end{center}
\caption{
Decomposition of the dilepton production rate.}
\label{fig:three}
\end{figure}

In order to understand these results in more detail,
we show the contribution of each term in Eq.~(\ref{eq:full_rate}) 
separately in Fig.~\ref{fig:three}.
In the figure, the rate coming from the pair annihilation of 
two normal modes (NN), two plasmino modes (PP),
and a normal and a plasmino modes (NP) are 
separately shown, together with those of the Landau damping between 
quasi-particles (LD) and processes including an $E$ mode with 
a normal (NE) and a plasmino (PE) modes.
From the figure, one finds that the two divergences at $\omega=\omega_1$
and $\omega_2$ come from the LD and PP rates, respectively.
As discussed in Sec.~\ref{sec:rate2}, these divergences 
come from van Hove singularity.
The photon self-energy Eq.~(\ref{eq:full_rate}) is 
inversely proportional to derivatives of the dispersion relations, 
$d\nu_-(p)/dp$ and $d\{\nu_+(p)-\nu_-(p)\}/dp$.
As shown in Figs. \ref{fig:one} and \ref{fig:add_disp}, 
each of $\nu_-(p)$ and $\nu_+(p)-\nu_-(p)$ has an extremum at nonzero $p$.
Their values at the extrema are $\nu_+(p)-\nu_-(p)=\omega_1$ and 
$2\nu_-(p)=\omega_2$.
At these points the derivatives vanish.
This leads to the divergences in 
the photon self-energy, and accordingly the dilepton production rate.

Figure~\ref{fig:three} also shows that each indivisual process 
is nonvanishing in a limited range of $\omega$.
The range can be read off from the corresponding functions plotted 
in Fig.~\ref{fig:add_disp}.
The NN and NP rates are nonvanishing for $\omega>2m_{\rm T}$.
On the other hand, the lower threshold of the PP rate, 
$\omega=\omega_2$, is slightly lower than $2m_{\rm T}$, because 
$\nu_-(p)$ has a minimum smaller than $m_{\rm T}$ at nonzero momentum.
The range of the Landau damping is also kinematically constrained 
to $\omega<\omega_1$.
The NE and PE rates are nonvanishing for 
$\omega>m_{\rm T}$ and $\omega<m_{\rm T}$, respectively.
The NE rate gives rise to a nonzero value for $\omega_1<\omega<\omega_2$.
To the dilepton pruduction rate without the vertex correction, 
only the NN, PP, and LD contribute and
the rate vanishes for $\omega_1<\omega<\omega_2$.

For large $\omega$, the rate is dominated by the NN.
This is a consequence of the fact that the quark propagator 
approaces the free quark one as $p$ becomes larger.
A glance at Fig.~\ref{fig:three} might give an impression that
the NE rate also survives for large $\omega$.
Although not shown in Fig.~\ref{fig:three}, however, the NP 
rate takes a negative value for $\omega\gtrsim2.4m_{\rm T}$,
and this term almost cancels out with the NE rate; 
see the discussion in Sec.~\ref{sec:rate3}.

Next, let us address the behavior of the production rate 
in the $\omega\to0$ limit.
The photon self-energy is identical with the 
electromagnetic current-current correlation function,
\begin{align}
  J_{ij}^R(\omega,\bm{p}) = 
  \int d^4x \mbox{e}^{i\omega t-i\bm{p}\cdot\bm{x}}
  \langle[j_i^{\mbox{\small EM}}(t,\bm{x}),j_j^{\small \mbox{EM}}(0,\bm{0})] \rangle\theta(t),
\end{align}
at the leading order in $\alpha$.
The low energy behavior of $J_{ij}^R(\omega,\bm{p})$ is related to 
electric conductivity $\sigma$ through the Kubo formula,
\begin{equation}
  \sigma=\frac16\lim_{\omega\to 0}\frac1\omega\sum_{i=1}^3
  J_{ii}^R(\omega,\bm{0}).
\end{equation}
Our result shows that
$\sum_{i=1}^3 J_{ii}^R(\omega,\bm{0})$
approaches zero faster than $\omega$ in the $\omega\to0$ limit,
and thus the electric conductivity vanishes.
Incorporation of the width of the quasi particle modes,
which is not included in the form of the quark propagator used 
in this study, may lead to nonzero $\sigma$.

\begin{figure}[htbp]
\begin{center}
\includegraphics{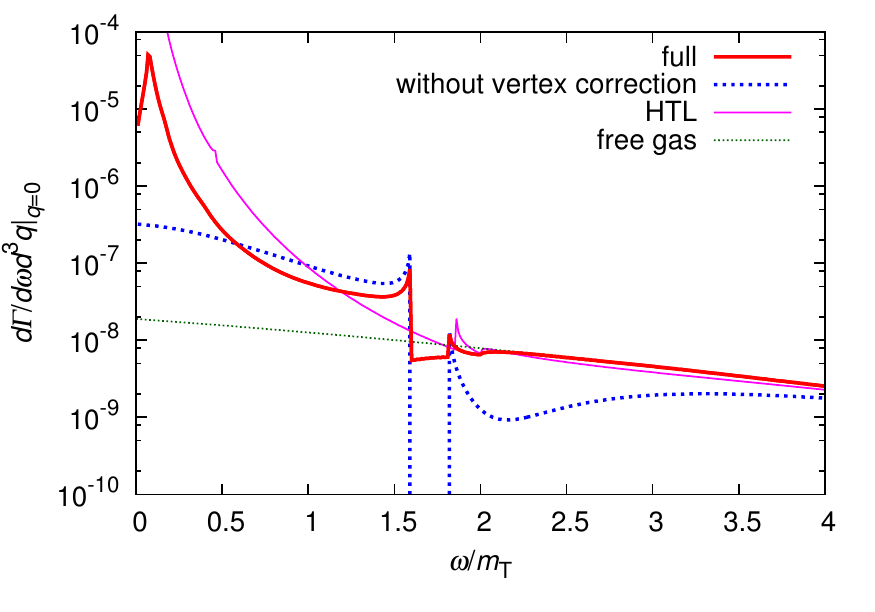}
\end{center}
\caption{Dilepton production rate from $3T_{\rm c}$ deconfined phase.}
\label{fig:five}
\end{figure}

In Fig.~\ref{fig:five}, we show the dilepton production rate 
at $T=3T_{\rm c}$.
One sees that the behavior is qualitatively the same as the 
result for $T=1.5T_{\rm c}$.
In particular, 
there exists an enhancement of the rate around 
$\omega\simeq 1.5 m_{\rm T}$ owing to van Hove singularity.
This indicates that the large enhancement of the dilepton 
production rate around $\omega\simeq m_{\rm T}$ is a general result
irrespective of $T$.
The gap between $\omega_1$ and $\omega_2$ is
narrower than that at $T=1.5T_{\rm c}$,
because of the change of the dispersion relations
$\nu_\pm(p)$ obtained on the lattice.
One also finds that the rate takes a finite value 
at $\omega=0$, while it diverges for $T=1.5T_{\rm c}$.
This limiting behavior may depend on the way of the extrapolation
of $\nu_\pm(p)$ to large momentum.

\section{\label{sec:level5} Summary}

In this study we have investigated the dilepton production 
rate using a quark propagator obtained on the lattice with 
a pole ansatz.
The Schwinger-Dyson equation for the photon self-energy
is solved with the lattice quark propagator and the photon-quark 
vertex satisfying the Ward-Takahashi identity.
The effect of the vertex correction is discussed by comparing 
the result with the calculation without vertex correction.
Our numerical result shows that the dilepton production rate 
with the lattice quark propagators is
larger by about one order or more
compared with that with the free quark ones
in the low invariant mass region.
Compared with the HTL result, there exists a significant 
enhancement around $\omega \simeq m_{\rm T} \simeq T$
owing to van Hove singularity.
This result is interesting since such a large enhancement in the 
deconfined medium near $T_{\rm c}$ may explain the excess of the 
dilepton yield observed at PHENIX \cite{a} in the low invariant mass region.
To understand the effect of the enhancement of the dilepton rate to 
the experimental result quantatively, the analysis with 
dynamical models describing the space-time evolution of the 
hot medium and integration of the dilepton production rate is needed,
which would be performed elsewhere.
The comparison with other non-perturbative approaches
such as Ref.~\cite{Gale:2014dfa}
will be an interesting future work.

The authors thank M.~Harada for a useful comment on the 
vertex funciton.
T.~K. is supported by Osaka University Cross-Boundary Innovation Program.
This work is supported in part by JSPS KAKENHI Grant
Numbers 23540307, 25800148, and 26400272.

\end{document}